\documentclass[twocolumn,letter]{jpsj2}
\usepackage{color}
\usepackage{ulem}
\usepackage{bm}

\def\runauthor{Online-Journal Subcommittee}

\title{%
Nonequilibrium relaxation study of the anisotropic antiferromagnetic Heisenberg model
on the triangular lattice
}

\author{%
Takahiro \textsc{Misawa}\thanks{E-mail:misawa@solis.t.u-tokyo.ac.jp} and
Yukitoshi \textsc{Motome}  
}

\inst{
{\it Department of Applied Physics, University of Tokyo,
7-3-1 Hongo, Bunkyo-ku, Tokyo, 113-8656, Japan}
}

\recdate{\today}

\abst{Effect of exchange anisotropy 
on the 
relaxation time of spin and vector chirality is studied for 
the antiferromagnetic classical Heisenberg model on the triangular lattice 
by using the nonequilibrium relaxation Monte Carlo method. 
We 
identify 
the Berezinskii-Kosterlitz-Thouless (BKT) transition and the chiral 
transition 
in a wide range of the anisotropy, 
even for very small anisotropy of $\sim 0.01$\%. 
As the anisotropy decreases, both the critical temperatures steeply decrease, 
while the BKT critical region becomes divergently wide. 
We elucidate 
a sharp ``V shape" of the phase diagram around the isotropic Heisenberg point, 
which suggests that the isotropic 
case is exceptionally singular and 
the associated $Z_2$ vortex transition 
will be
isolated from 
the BKT and chiral transitions.
We 
discuss the relevance of our results to 
peculiar behavior of the
spin relaxation time observed experimentally 
in triangular antiferromagnets.
}

\kword{geometrical frustration, triangular lattice, Heisenberg model,
nonequilibrium relaxation, BKT transition, chiral transition, $Z_2$ vortex transition
}

\begin{document}
\maketitle

Antiferromagnet 
on the two-dimensional triangular lattice 
has been intensively studied 
as one of the most fundamental models for the geometrically frustrated systems~\cite{Diep}.
For the isotropic 
Heisenberg model with nearest-neighbor interactions, 
it is 
believed that 
the ground state of the system exhibits a three-sublattice 
120$^{\circ}$ 
long-range order~\cite{Capriotti_1}, 
whereas 
the magnetic 
ordering is no longer retained 
against thermal fluctuations~\cite{Mermin1966}.
Nevertheless, an interesting possibility was proposed 
by Kawamura and Miyashita~\cite{KM_1,KM_2}, that is,  
an unconventional topological transition at a finite temperature ($T$) 
--- $Z_{2}$ vortex transition. 
From the symmetry point of view, 
the $Z_2$ vortex transition is different from 
the conventional 
Berezinskii-Kosterlitz-Thouless (BKT) transition 
which occurs in the presence of anisotropy~\cite{Berezinskii1970,Kosterlitz1973}. 
The relation between these two topological transitions, however, 
is not fully understood yet. 
In particular, it is still unclear 
how the system behaves in the region of vanishing anisotropy. 

Experimentally, several materials with triangular layered structure have been studied, 
and recently, the $Z_2$ vortex 
attracts renewed interests
for understanding of their peculiar properties. 
One of the peculiar properties is anomalous enhancement of 
the spin relaxation time. 
Critical divergence of 
the relaxation time 
is observed 
in an anomalously wide range of $T$ 
in many compounds, 
such as 
$A$CrO$_2$ ($A$=Li,H,Na)~\cite{Ajiro,LiCrO2_2,NaCrO2,Hemmida2009},
Li$_{7}$RuO$_{6}$~\cite{Li7RuO6}, 
and 
NiGa$_2$S$_4$~\cite{NiGa2S4_2,Yamaguchi2008}. 
The critical behaviors 
are often 
argued to be a fingerprint of the $Z_2$ vortex transition.  
The $Z_2$ vortex, however, is a 
topological object 
specific to 
spin-rotational-invariant 
systems,
and hence, 
it is not trivial whether its influence is observed in 
real compounds 
in which 
anisotropy 
exists. 

In this letter, to shed light on the origin of the
anomalous critical behavior 
and its relation to the $Z_2$ vortex transition, 
we directly calculate the relaxation time in 
the antiferromagnetic Heisenberg model with classical spins on the triangular lattice.
We 
focus on how the anisotropy in exchange interactions
affects the behavior of 
relaxation time. 
We 
determine the finite-$T$ phase diagram precisely 
by varying the anisotropy, 
and uncover the exceptionally singular nature of the isotropic Heisenberg case. 

Our model Hamiltonian is defined in the form
\begin{equation}
H=J\sum_{\langle ij\rangle}(S_{i}^{x}S_{j}^{x}+S_{i}^{y}S_{j}^{y}+\lambda S_{i}^{z}S_{j}^{z}),
\label{eq:H}
\end{equation}
where $J$ is the antiferromagnetic exchange interaction, 
and 
$\bm{S}_i = (S_{i}^{x},S_{i}^{y},S_{i}^{z})$ is a vector representing the classical spin at site $i$
(we normalize as $|\bm{S}_i| = 1$); 
the summation $\langle ij\rangle$ runs over the nearest-neighbor
bonds of the triangular lattice.
We introduce here the exchange anisotropy $\lambda$. 
For the $XY$ anisotropy ($\lambda<1$),
it is known that 
both BKT and chiral 
transitions occur 
at different but very close temperatures~\cite{MiyashitaShiba,Capriotti_2}.
On the other hand, for the Ising anisotropy 
($\lambda>1$), it is known that
two different BKT transitions occur separately
for the longitudinal $S^{z}$ and 
the transverse $(S^{x},S^{y})$ components~\cite{MiyashitaKawamura,StephanSouthern_1}.
It is, however, still controversial how these four transitions behave 
as the system approaches the isotropic Heisenberg point $\lambda=1$. 
We will discuss this issue later.

\begin{figure}[t]
	\begin{center}
		\includegraphics[width=8cm,clip]{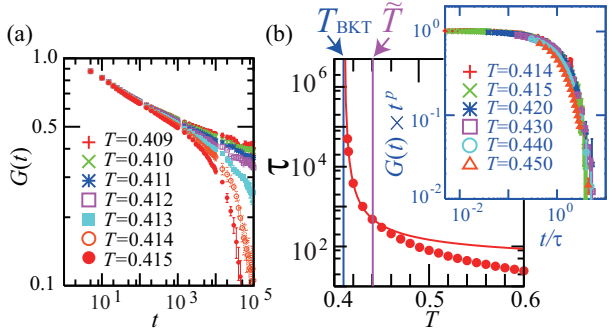}   
	\end{center}
\caption{(Color online)~(a)~Dynamical spin correlation $G(t)$ 
as a function of the Monte Carlo step $t$ 
for the model (\ref{eq:H}) in the $XY$ limit $\lambda=0$.
(b)~Spin relaxation time $\tau$ as a function of temperature $T$ at $\lambda=0$.
The curve shows a fit by the BKT scaling $\tau = a\exp[b/(T-T_{\rm BKT})^{1/2}]$. 
$T_{\rm BKT}$ and $\tilde{T}$ are the 
estimates of the
BKT transition temperature and 
the onset of the BKT critical region, respectively. See the text for details. 
The inset demonstrates the scaling plot for {$G(t)$}. Below $\tilde{T}$,
all the data are scaled well to a single universal function 
{(we discard the data for $t \le 100$)}.
}
\label{fig:Fig1}
\end{figure}%

We calculate the relaxation time of the model (\ref{eq:H}) 
by using the nonequilibrium relaxation (NER) method~\cite{NER_1}. 
In this method, 
the relaxation time is directly computed 
by analyzing the relaxation process from an initial 
ordered
state 
in terms of the Monte Carlo (MC) dynamics. 
We typically perform the relaxation up to 10$^5$ MC steps 
by using the standard Metropolis local update 
for the system size $N_{\rm s}=L\times L$
up to $L=4002$ under the periodic boundary conditions. 
We confirm that the finite-size effect is negligibly small. 
The results are averaged over eight independent MC runs. 
We choose the initial 
state 
to be a three-sublattice $120^{\circ}$ state; 
spins are set to be in the $xy$ plane 
for 
$\lambda\leq 1$,
while they are 
in 
the $xz$ plane with aligning one of three spins to the $z$ direction 
for 
$\lambda>1$.
The 
ground state is slightly
different from the $120^{\circ}$ 
state in the Ising case $\lambda>1$, 
but
this deviation does not affect the long-time behavior of relaxation process. 
We set $J=1$ and the Boltzmann constant $k_{\rm B}=1$. 

In Fig.~\ref{fig:Fig1}, 
we demonstrate how the NER method works 
in the $XY$ limit ($\lambda=0$), as an example. 
We calculate the dynamical spin correlation function $G(t)$ defined as
\begin{equation}
G(t)=\frac{1}{N_{\rm s}}\sum_{i} 
\langle 
\bm{S}_{i}(t)\cdot\bm{S}_{i}(0)
\rangle,
\label{eq:G(t)}
\end{equation}
where 
$\bm{S}_{i}(t)$
denotes the spin configuration at site $i$
and MC step $t$.
As shown in Fig.~\ref{fig:Fig1}(a), 
$G(t)$ changes 
from an exponential decay 
in the high-$T$ paramagnetic phase 
to a power-law decay 
in the low-$T$ BKT phase. 
The BKT transition temperature 
$T_{\rm BKT}$ is determined by the divergence of the relaxation time $\tau$ 
estimated from the high-$T$ exponential behavior, $G(t)\sim \exp(-t/\tau)$. 
%
We employ the scaling analysis 
by using $G(t)=\tau^{-p}f(t/\tau)$~\cite{NER_2}, 
which enables us to estimate 
$\tau$ up to $\sim 10^{5}$.
The results are plotted in Fig.~\ref{fig:Fig1}(b). 
The divergent behavior of $\tau$ at low $T$ 
is well fitted by the BKT scaling 
$\tau=g(T)=a\exp[b/(T-T_{\rm BKT})^{1/2}]$. 
Here we 
choose the $T$ range 
for the fit 
by monitoring 
the weighted residual 
defined by 
$R_{\rm w} \equiv \frac{1}{N_T} \sum_{i=1}^{N_T} [ \{\tau_i - g(T_i) \}/g(T_i) ]^2$, 
where $N_T$ is the number of the data $\{ T_i \}$: 
We fit the range of low-$T$ data which gives $R_{\rm w}$ less than $0.002$. 
The fit 
gives an estimate of the transition temperature $T_{\rm BKT}=0.409(1)$, 
which is consistent with the previous estimate\cite{Capriotti_2}. 
At the same time, the fitting procedure defines 
$\tilde{T}=0.440(10)$, 
%
below which 
$\tau$ follows the
BKT scaling. 
The 
region 
$T_{\rm BKT} < T < \tilde{T}$ 
represents the BKT critical region
in which {$G(t)$} obeys the universal {behavior} [see the inset of Fig. 1(b)].
We confirm a similar universal scaling for all the following data.

\begin{figure}[t]
	\begin{center}
		\includegraphics[width=8cm,clip]{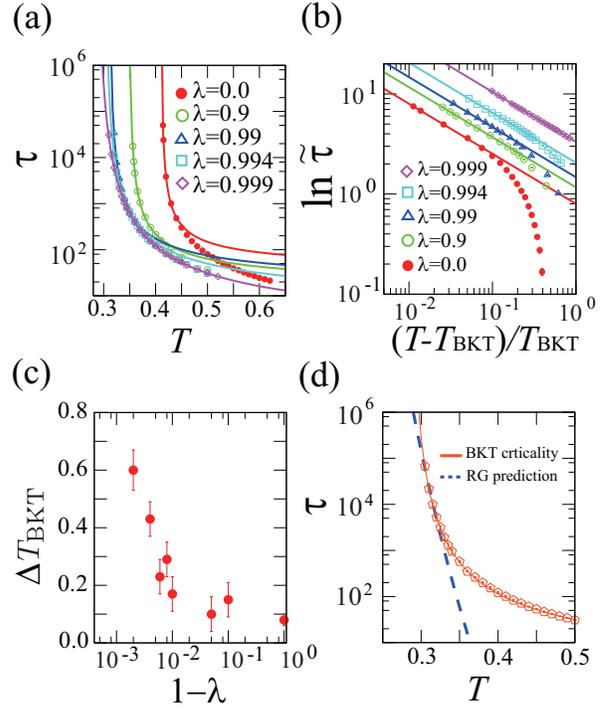}   
	\end{center}
\caption{(Color online) 
(a)~Critical behaviors of $\tau$ for various values of the anisotropy $\lambda$. 
(b)~BKT scaling plot for the normalized relaxation time $\tilde{\tau}$. 
The lines show the BKT scaling fit. 
(c) Relative width of the BKT critical region 
as a function of the anisotropy. 
(d) 
$T$ dependence of $\tau$ at the Heisenberg point $\lambda=1$. 
The 
curve shows the BKT fit, and 
the dashed 
line represents the fitting 
on the basis of 
the renormalization-group (RG) analysis 
for the non-linear $\sigma$ model, 
$\tau=C_{t}[(T/B)^{x}\exp(B/T)]^{z}$~\cite{Azaria}.
}
\label{fig:Fig2}
\end{figure}

We study the critical behavior of $\tau$ in this way 
for various values of the anisotropy $\lambda$. 
The results for $\lambda < 1$ are 
shown in Fig.~\ref{fig:Fig2}(a). 
As approaching the Heisenberg case with $\lambda \to 1$, 
$T_{\rm BKT}$ 
decreases monotonically, 
while $\tilde{T}$ first decreases but increases for $\lambda > 0.9$:
As a result, 
the width of the BKT critical region becomes 
wider as $\lambda \to 1$.
This is more clearly observed in Fig.~\ref{fig:Fig2}(b), 
which plots 
the normalized relaxation time 
$\tilde{\tau} = \tau/a$ 
on the basis of the BKT scaling  
$\tau = a\exp[b/(T-T_{\rm BKT})^{1/2}]$. 
In fact, 
as shown in Fig.~\ref{fig:Fig2}(c), 
the relative width of the critical region, 
$\Delta T_{\rm BKT} = (\tilde{T} - T_{\rm BKT}) / T_{\rm BKT}$, 
appears to diverge 
logarithmically or more strongly 
with decreasing the anisotropy. 

As anticipated 
from the 
diverging $\Delta T_{\rm BKT}$, 
$\tau$ for the isotropic Heisenberg case can be fitted by the BKT scaling 
in the entire range of $T$ calculated, 
as shown in Fig.~\ref{fig:Fig2}(d). 
The fitting naively suggests that $\tau$ diverges 
at 
$T^{*}=0.282(4)$. 
Similar behavior was seen in the spin correlation length, 
for which it was argued that
a crossover takes place 
from the BKT behavior to another $T$ dependence 
and the correlation length does not diverge except for $T=0$~\cite{StephanYoung,Wintel}. 
Our data are consistent with such analyses as shown in Fig.~\ref{fig:Fig2}(d). 
We return to this point later. 
Besides the crossover, 
the crucial observation here is that the isotropic case $\lambda=1$ looks quite singular 
since the apparent BKT critical 
behavior is 
observed in the divergently wide range of $T$.

\begin{figure}[t]
	\begin{center}
		\includegraphics[width=8.5cm,clip]{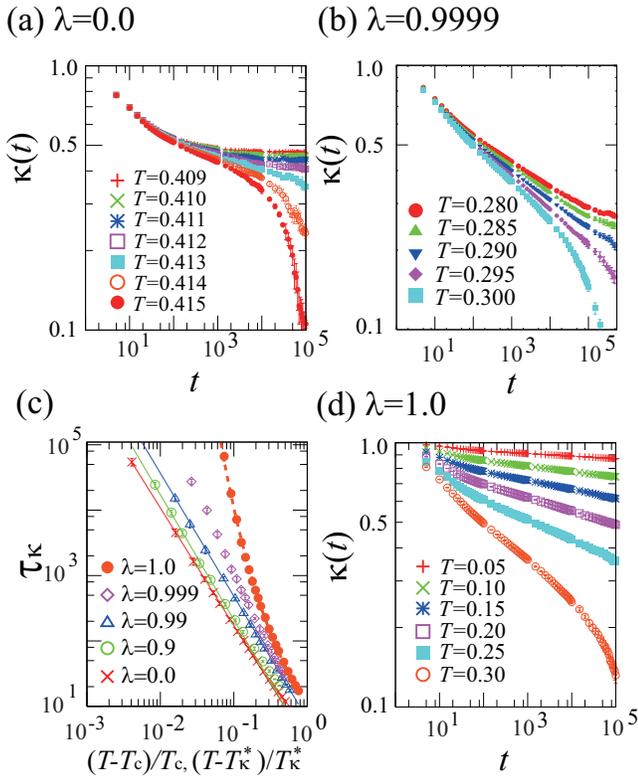}   
	\end{center}
\caption{(Color online)~
Relaxation of the vector chiral order parameter 
(a) in the $XY$ limit ($\lambda=0.0$) and
(b) 
very close to the Heisenberg point ($\lambda=0.9999$).
(c)~Scaling plot of 
the relaxation time of the chirality, 
$\tau_{\kappa}$. 
Solid lines 
for $\lambda \le 0.99$
represent the power-law fit $\tau_{\kappa}\propto (T-T_{c})^{-z\nu}$.
The dashed curve for the Heisenberg case $\lambda=1$ is 
the fit to 
the BKT criticality, i.e.,
$\tau_{\kappa}\propto \exp[b/(T-T_\kappa^{*})^{1/2}]$.
(d)~Relaxation 
in the Heisenberg case. 
}
\label{fig:Fig3}
\end{figure}%

Let us further discuss the singular behavior as $\lambda \to 1$ 
from the viewpoint of the vector spin chirality. 
We study 
the dynamical chiral correlation function 
defined as
\begin{equation}
\kappa(t)=\frac{1}{2N_{\rm s}}\sum_{\bm{R}_{i}}
\langle \kappa^{z}_{\bm{R}_{i}}(t) 
\,
\kappa^{z}_{\bm{R}_{i}}(0)
\rangle,
\end{equation}
where the summation $\bm{R}_{i}$ 
runs over all the unit triangles, and 
$\kappa^{z}$ is the $z$ component of the vector chirality 
$
\kappa^{z}=\frac{2}{3\sqrt{3}}(\bm{S}_{1}\times \bm{S}_{2}+\bm{S}_{2}\times \bm{S}_{3}
+\bm{S}_{3}\times \bm{S}_{1})^{z} 
$
for each triangle 
of three spins 
$\bm{S}_1$, $\bm{S}_2$, and $\bm{S}_3$.
The vector chirality exhibits a true long-range order in the anisotropic cases, 
and therefore, we expect that 
$\kappa(t)$ decays 
exponentially above 
a chiral transition temperature $T_{\rm c}$, 
while it approaches a nonzero constant 
below $T_{\rm c}$: 
At $T=T_{\rm c}$, $\kappa(t)$ shows a power-law decay. 
This is indeed the case as demonstrated 
in Figs.~\ref{fig:Fig3}(a) and \ref{fig:Fig3}(b) 
for $\lambda=0.0$ and $0.9999$, respectively.
The results give the estimates 
$T_{\rm c}=0.413(1)$ for $\lambda=0.0$~\cite{Capriotti_2} and 
$T_{\rm c}=0.290(5)$ for $\lambda=0.9999$. 
It is noteworthy that the 
chiral transition 
is clearly discernible even 
for very small anisotropy of 0.01\% ($\lambda=0.9999$).

Similar to the analysis of $\tau$ in Fig.~\ref{fig:Fig2}, 
we examine the behavior of the relaxation time of the chirality, $\tau_\kappa$, 
by varying the anisotropy $\lambda$. 
The results for $\lambda<1$ are summarized in Fig.~\ref{fig:Fig3}(c). 
For $\lambda < 1$, $\tau_\kappa$ shows a power-law divergence 
$\tau_\kappa \propto (T-T_{\rm c})^{-z\nu}$, 
with the same exponent $z\nu \simeq 1.9$.
At $\lambda=0.999$ the data show crossover from a BKT-like behavior at high $T$ to the power-law divergence near $T_{\rm c}$.
In the isotropic case $\lambda=1$, 
$\tau_\kappa$ shows a stronger divergence than the power law and 
is well fitted by the BKT scaling $\tau_{\kappa}\propto \exp[b/(T-T_\kappa^{*})^{1/2}]$ 
with $T_\kappa^{*} = 0.284(3)$. 
The relaxation process at $\lambda=1$ is 
shown in Fig.~\ref{fig:Fig3}(d); 
$\kappa(t)$ exhibits a power-law decay even at much lower $T$ than $T_\kappa^*$. 
The 
results 
seemingly suggest 
a BKT-type transition 
in the vector chiral degree of freedom at $T_\kappa^*$, 
as seen at $T^*$ in the spin sector [Fig.~\ref{fig:Fig2}(d)]. 
These behaviors in the vector chirality 
also illuminate a singularity of the isotropic Heisenberg case. 
We will 
discuss $T^{*}$ and $T_{\kappa}^{*}$ 
in comparison with the $Z_2$ vortex transition temperature later.

Collecting the data of the relaxation 
of spin and vector chirality,
we map out the finite-$T$ phase diagram as a function of the anisotropy $\lambda$. 
We also study 
the Ising anisotropic cases ($\lambda>1$), 
where two different BKT transitions take place; 
one is for the $z$ component of spin $S^z$, and 
the other is for the $xy$ components $S^x$, $S^y$~\cite{MiyashitaShiba}. 
The latter transition 
accompanies a quasi long-range ordering of the vector chirality, 
and occurs at a lower $T$ than the former~\cite{StephanSouthern_1}. 
The two BKT transition temperatures are determined from 
the dynamical spin correlation function for the corresponding spin component [cf. Eq.~(\ref{eq:G(t)})]. 

\begin{figure}[t]
	\begin{center}
		\includegraphics[width=8.5cm,clip]{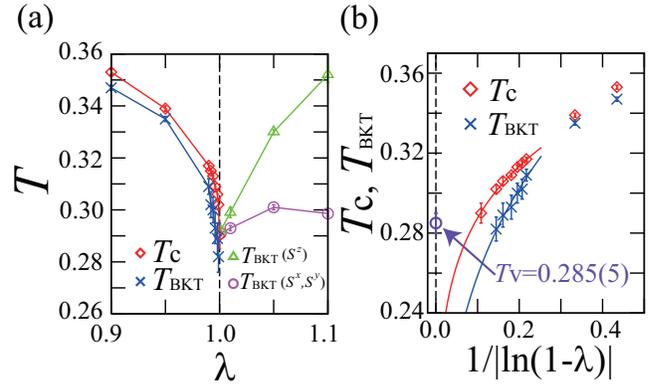}   
	\end{center}
\caption{(Color online)~(a) Phase diagram for the anisotropic Heisenberg model 
(\ref{eq:H}) 
determined by the NER method.
For the $XY$ anisotropy ($\lambda<1$), 
the chiral 
and 
BKT transition 
temperatures are shown by diamonds and crosses, respectively. 
For the Ising anisotropy ($\lambda>1$),
two BKT transition temperatures 
as to $S^z$ and $(S^x, S^y)$ components are plotted 
by triangles and circles, respectively.
The lines are guides for the eye. 
(b) 
Chiral and BKT transition temperatures 
as a function of 
$1/|\ln(1-\lambda)|$ for $\lambda < 1$. 
The data are fitted by $1/|\ln(1-\lambda)|^{\alpha}$. 
For comparison, a recent estimate of $Z_{2}$ vortex transition temperature $T_v$
is shown~\cite{Kawamura2009}. 
}
\label{fig:Fig4}
\end{figure}%

Figure~\ref{fig:Fig4}(a) summarizes our phase diagram 
around the isotropic Heisenberg point $\lambda \sim 1$. 
In 
the $XY$ anisotropic region $\lambda<1$, 
the chiral transition always occurs at a slightly higher $T$ than the BKT transition\cite{Capriotti_2}. 
Both two transition temperatures 
decrease 
more rapidly as $\lambda \to 1$. 
Similar decrease 
is observed 
in the two BKT transitions 
when 
approaching from the Ising anisotropic case $\lambda>1$. 
Especially, the BKT transition temperature for the $xy$ components 
shows 
a nonmonotonic $\lambda$ dependence. 
Consequently, the phase diagram exhibits a sharp ``V shape", 
which illuminates the singularity of the Heisenberg case $\lambda = 1$. 
To our knowledge, this peculiar form 
in the very vicinity of $\lambda=1$ 
has not been elucidated before~\cite{StephanSouthern_2}.

The question is the fate of the transition temperatures as $\lambda \to 1$, 
in particular, 
their relation to the $Z_{2}$ vortex transition predicted for 
$\lambda=1$.
Figure~\ref{fig:Fig4}(b) shows the asymptotic behavior of the chiral and BKT transition temperatures 
in comparison with a recent estimate of $Z_2$ vortex transition temperature $T_v$~\cite{Kawamura2009}.
We plot the data as a function of $1/|\ln{(1-\lambda)}|$, 
with considering 
an analytical argument 
for the square lattice model  
which predicts $T_{\rm BKT} \propto 1/|\ln{(1-\lambda)}|$ 
on the basis of independent vortex pair picture~\cite{Hikami}. 
Surprisingly, both $T_{\rm c}$ and $T_{\rm BKT}$ decrease faster than $\propto 1/|\ln (1-\lambda)|$, 
suggesting that they will be well below $T_v$ and finally approach zero as $\lambda \to 1$. 
{(The fitting in the figure shows $1/|\ln (1-\lambda)|^\alpha$ with $\alpha < 1$ as a guide.)}
From these observations, we conclude that 
the phase boundaries of the chiral and BKT transitions 
show the highly-singular ``V-shape" around the isotropic point $\lambda=1$, 
and 
the $Z_2$ vortex transition is isolated from the conventional phase transitions. 

We note that the $Z_2$ vortex transition temperature $T_v = 0.285(5)$~\cite{Kawamura2009} 
coincides with the apparent `transition temperatures' $T^* = 0.282(4)$ estimated from 
the BKT fitting of $\tau$ [Fig.~\ref{fig:Fig2}(d)] and 
$T_\kappa^* = 0.284(3)$ similarly obtained for $\tau_\kappa$ [Fig.~\ref{fig:Fig3}(c)]. 
Since the relaxation time $\tau$ and $\tau_\kappa$ exceed $10^6$, 
it is hard to trace a crossover from the BKT scaling to 
another behavior, if any, 
within the accessible system size [see Fig.~\ref{fig:Fig2}(d)].
Although dynamics of $Z_{2}$ vortices may cause these diverging
behaviors~\cite{KM_2}, their relation is not clear. 
Nonetheless, as discussed above, since we expect a finite spin correlation length 
for $T>0$ at $\lambda={1}$, 
it is natural to consider $T_{\rm BKT} \to 0$ as $\lambda \to 1$, not $T_{\rm BKT} \to T^*$. 
The situation is not so clear for the vector chirality, but 
it is plausible that $T_{\rm c}$ also goes to zero as $\lambda \to 1$, 
because $T_{\rm c}$ 
coincides with $T_{\rm BKT}$ for $S^x, S^y$ 
in the Ising case $\lambda>1$, which should go to zero. 
The asymptotic behaviors in Fig.~\ref{fig:Fig4}(b) support our consideration. 

Finally, let us discuss the relevance of our results to experiments. 
The peculiar ``V shape" phase diagram in Fig.~\ref{fig:Fig4} indicates that 
the anisotropy is a relevant perturbation to the isotropic Heisenberg point $\lambda=1$ 
in the sense not only that it 
triggers the finite-$T$ transitions 
but also that the induced transition temperatures grow in a very singular fashion against the anisotropy.
Therefore, 
in the triangular antiferromagnets, 
the isotropic Heisenberg case is special, rather isolated; 
the pristine property of the isotropic point including the $Z_2$ vortex transition is hardly accessible 
in real materials in which anisotropy exists inevitably. 
For example, 
NaCrO$_2$ is known to have 
a small but finite anisotropy in the $g$ value about 0.25\%~\cite{Elliston1975},
and NiGa$_2$S$_4$ has a rather substantial anisotropy of $\sim 3$\%~\cite{Yamaguchi2008}. 
Although there might be a difference 
between the exchange anisotropy and the single-ion anisotropy~\cite{Melchy2009}, 
our results strongly suggest that in real materials 
the conventional BKT transition 
dominates the critical behavior of two-dimensional spin fluctuations, 
instead of the unconventional $Z_2$ vortex transition. 
The anomalous enhancement of the spin relaxation time will be understood 
by the divergently-enlarged BKT critical region near the isotropic case. 

A potential ``smoking gun" for the relevance of anisotropy is 
critical behavior of the relaxation time of the vector chirality. 
As shown in Fig.~\ref{fig:Fig3}(c), 
the power-law criticality is observed in the presence of the anisotropy, 
whereas the BKT-type divergence dominates at the isotropic point. 
It is difficult to probe the chirality experimentally, 
but we note that recently the chiral phase transition 
was detected by 
polarized neutron scattering~\cite{Plakhty2006}. 
The development along this direction is highly desired. 

In summary, by using the nonequilibrium relaxation Monte Carlo method, 
we have studied the relaxation time of spin and vector chirality 
in the anisotropic classical Heisenberg model on the triangular lattice. 
For the spin relaxation time, we have revealed that the BKT critical region 
becomes divergently wide 
as the anisotropy decreases. 
For the vector chirality, 
the relaxation time 
exhibits a power-law divergence, 
whereas at the isotropic Heisenberg point, it 
shows an apparent BKT criticality in a wide range of temperatures.
We have also obtained the precise phase diagram 
with its singular ``V shape" around the isotropic Heisenberg point, 
which uncovers 
the singular nature of the isotropic point. 
Our findings will be important for understanding of the puzzling experimental results 
in the triangular antiferromagnets 
as well as of unsettled theoretical issue on the relation between 
the conventional transitions and the $Z_2$ vortex transition. 

\acknowledgements{
The authors thank 
Hikaru Kawamura, 
Clare Lhuillier, 
Seiji Miyashita, 
Yohsuke Murase,
Tsuyoshi Okubo,  
Youhei Yamaji, 
and Mike Zhitomirsky
for fruitful discussions. 
YM also acknowledges the hospitality of KITP Santa barbara 
where the early stage of this work was completed. 
This work was supported by Grant-in-Aid for Scientific Research (No. 19052008), 
Global COE Program ``the Physical Sciences Frontier", and 
the Next Generation Super Computing Project, Nanoscience Program,
from 
MEXT, Japan. 
}

\end{document}